\documentclass[conference]{IEEEtran}

\usepackage{graphicx}
\usepackage{tabularx}

\usepackage[T1]{fontenc}
\usepackage[utf8]{inputenc}

\usepackage{csquotes}
\usepackage{nicefrac}
\usepackage{afterpage}
\usepackage{needspace}
\usepackage[textsize=scriptsize]{todonotes}
\usepackage{booktabs}
\usepackage{xcolor}
\usepackage{tikz}
\usepackage{quantikz}
\usetikzlibrary{arrows,positioning} 
\usetikzlibrary{shapes.geometric, positioning}
\usepackage{flushend}
\usepackage{soul}
\usepackage{amssymb}
\usepackage{csvsimple}
\usepackage{amsthm}
\usepackage{lipsum}
\usepackage{varwidth}
\usepackage{utfsym}
\usepackage{fontawesome5}
\usepackage{multicol}
\usepackage{listings}
\usepackage{adjustbox}

\robustify\bfseries

\makeatletter
\let\MYcaption\@makecaption
\makeatother
\usepackage[font=footnotesize]{subcaption}
\makeatletter
\let\@makecaption\MYcaption
\makeatother

\usepackage{yquant}
\usepackage{pgfplots}

\usepackage[style=ieee, maxnames=6, minnames=1, date=year, doi=false,isbn=false,backend=biber]{biblatex}
\usepackage[detect-all=true]{siunitx}
\usepackage{etoolbox}
\robustify\bfseries

\usetikzlibrary{fit, quotes}
\usetikzlibrary{shapes.geometric, positioning}
\usepackage{bbm}

\newtheorem{example}{Example}

\usetikzlibrary{automata,arrows}

\newcommand{\squigglyunderline}[1]{%
  \tikz[baseline=(X.base)]{
    \node[inner sep=0pt] (X) {#1};
    \draw[line width=0.5pt, draw=red, decorate,decoration={snake,amplitude=0.2mm,segment length=2mm}] ([yshift=-0.1mm] X.south west) -- ([yshift=-0.1mm] X.south east);
  }%
}

\newcommand{\warningtriangle}{%
\begin{tikzpicture}[baseline=-0.2ex]
  \node[regular polygon, regular polygon sides=3, fill=yellow,scale=0.8] (triangle) at (0,0) {};
  \node[inner sep=0pt, anchor=center] at (triangle.center) {\fontsize{4pt}{5pt}\selectfont !};
\end{tikzpicture}
}

\newcommand{\graybox}[1]{%
  \tikz[baseline=-0.4em]{
    \node[fill=gray!30, text width=4cm, align=left, inner sep=1.5pt] {\fontsize{5pt}{6pt}\selectfont \warningtriangle #1};
  }%
}

\AtEveryBibitem{
  \clearname{editor}%
  \clearfield{series}%
  \clearfield{isbn}%
  \clearfield{issn}%
  \clearfield{volume}%
  \clearfield{number}%
  \clearfield{pages}%
  \clearfield{doi}%
}

\lstdefinelanguage{mqt-qasm}{
  keywords={include, qreg, creg, gate, measure, reset, barrier, if, opaque},
  keywordstyle=\color{blue}\bfseries,
  ndkeywords={x, h, cccz, ccz, oracle, diffusion},
  ndkeywordstyle=\color{darkgray}\bfseries,
  identifierstyle=\color{black},
  sensitive=false,
  comment=[l]{//},
  morecomment=[s]{/*}{*/},
  commentstyle=\color{green}\ttfamily,
  stringstyle=\color{red}\ttfamily,
  morestring=[b]",
  morestring=[b]',
  basicstyle=\ttfamily\footnotesize,
  numbers=left,
  numberstyle=\tiny\color{gray},
  stepnumber=1,
  numbersep=5pt,
  showspaces=false,
  showstringspaces=false,
  showtabs=false,
  tabsize=2,
  breaklines=true,
  breakatwhitespace=false,
  captionpos=b,
  frame=single,
  alsoletter=-,
  lineskip=0pt,
}

\lstdefinestyle{qasm}{
    commentstyle=\color{green},
    keywordstyle=\color{blue},
    numberstyle=\tiny\color{gray},
    stringstyle=\color{red},
    basicstyle=\ttfamily\footnotesize,
    breakatwhitespace=false,         %
    breaklines=true,                 %
    captionpos=b,                    %
    keepspaces=true,                 %
    numbers=left,                    %
    numbersep=5pt,                   %
    showspaces=false,                %
    showstringspaces=false,          %
    showtabs=false,                  %
    tabsize=2,                       %
    frame=single,                    %
}
\newif\ifdoubleblind
\doubleblindfalse

\newcommand{\redacted}[1]{\ifdoubleblind[redacted for double-blind review]\else#1\fi}
\newcommand{\redactedInstead}[2]{\ifdoubleblind#2\else#1\fi}
\redactedInstead{}{}

\usepackage{hyperref}

\hypersetup{
	pdftitle={A Framework for Debugging Quantum Programs},   %
	pdfsubject={Design Automation Conference 2025}, %
	pdfauthor={\redactedInstead{Damian Rovara, Lukas Burgholzer, Robert Wille}{}}
}

\begin{document}

\title{A Framework for Debugging Quantum Programs}

\author{
  \redactedInstead{
	\IEEEauthorblockN{Damian Rovara\IEEEauthorrefmark{1}\hspace*{1.5cm}Lukas Burgholzer\IEEEauthorrefmark{1}\hspace*{1.5cm}Robert Wille\IEEEauthorrefmark{1}\IEEEauthorrefmark{2}}
	\IEEEauthorblockA{\IEEEauthorrefmark{1}Chair for Design Automation, Technical University of Munich, Germany}
	\IEEEauthorblockA{\IEEEauthorrefmark{2}Software Competence Center Hagenberg GmbH (SCCH), Austria}
	\IEEEauthorblockA{\href{mailto:damian.rovara@tum.de}{damian.rovara@tum.de}\hspace{1.5cm}\href{mailto:lukas.burgholzer@tum.de}{lukas.burgholzer@tum.de}\hspace{1.5cm} \href{mailto:robert.wille@tum.de}{robert.wille@tum.de}\\
	\url{https://www.cda.cit.tum.de/research/quantum}}
  \vspace{-0.75cm}
  }{\vspace{1cm}}
}

\maketitle

\begin{abstract}
  Recent advancements in quantum computing software are gradually increasing the scope and size of quantum programs being developed. At the same time, however, these larger programs provide more possibilities for functional errors that are harder to detect and resolve. Meanwhile, debugging tools that could aid developers in resolving these errors are still barely existent and far from what we take for granted in classical design automation and software engineering.
  As a result, even if one manages to identify the incorrect behavior of a developed quantum program, detecting and resolving the underlying errors in the program remains a time-consuming and tedious task. Moreover, the exponential growth of the state space in quantum programs makes the efficient manual investigation of errors radically difficult even for respectively simple algorithms, and almost impossible as the number of qubits increases. To address this problem, this work proposes a debugging framework, available as an open-source implementation at \redacted{\href{https://github.com/cda-tum/mqt-debugger}{https://github.com/cda-tum/mqt-debugger}}. It assists developers in debugging errors in quantum programs, allowing them to efficiently identify the existence of errors \emph{and} diagnose their causes. Users are given the ability to place assertions in the code that test for the correctness of a given algorithm and are evaluated using classical simulations of the underlying quantum program. Once an assertion fails, the proposed framework employs different diagnostic methods to point towards possible error causes. This way, the debugging workload for quantum programs is drastically reduced.
\end{abstract}

\section{Introduction}\label{sec:intro}

Due to the complexity of quantum computing, fully understanding larger quantum algorithms and the impact of all individual operations within them can be a challenging task, even for domain experts. Because of this, identifying errors and finding their underlying root causes often requires a large effort. Operations rarely commute, the exponential state space requires developers to keep track of a large number of amplitudes, and consequences of incorrect operations may sometimes only materialize at a much later step of execution.

Because of this, strategies to assist developers in checking and debugging quantum programs are becoming a rapidly growing area of research~\cite{dimatteo2024, li2014, huang2019a, metwalli2022, miranskyy2021, garciadelabarrera2023, metwalli2024}. Here, proven methods from classical debugging, such as interactive debuggers~\cite{quantag, jiang2023} and assertions~\cite{liu2019, haner2020, yu2021, ying2022, bauer-marquart2023, huang2019, li2020, liu2020, liu2021, witharana2023}, are employed to find and resolve errors in quantum programs efficiently. However, they are still limited in their scope and usability compared to their classical counterparts. More precisely, while the state of the art in quantum program debugging offers some promising methods to check quantum algorithms for their correctness, they still leave the task of identifying the underlying error to the developers---requiring further intensive work and extensive domain knowledge to resolve the problem. 

To address this issue, this work proposes a comprehensive framework for debugging quantum programs, providing the following features:
\begin{itemize}
    \item The framework employs \emph{assertions} (as introduced in~\cite{liu2019, haner2020, yu2021, ying2022, bauer-marquart2023, huang2019, li2020, liu2020, liu2021, witharana2023}) to check for the correctness of a given quantum program. 
    \item The framework employs \emph{classical simulation} to execute the quantum program and check the assertions. This way, the framework supports detailed insight into the system's quantum state during execution, circumventing limitations encountered on physical quantum devices, such as the \emph{no-cloning theorem}~\cite{wootters1982}, that would otherwise greatly limit the possibilities of interaction with the current state.
    \item Finally, the framework employs automated \emph{diagnosis techniques} to point towards possible error causes. Through continuous access to the full quantum state, the framework is able to analyze the program and search for anomalies that may be related to failing assertions. At the same time, the framework inspects dependencies of individual instructions and interactions of specific qubits, allowing it to reason about causes for misconfiguration by analyzing the corresponding dependency graph.
\end{itemize}

Through its modular nature, the proposed framework can support a variety of different simulation backends, allowing it to take advantage of various optimized simulators to debug larger quantum circuits. While there is an inevitable limit to the sizes of circuits that can be simulated, this approach can be employed to test circuits on a smaller scale and extrapolate their correctness for larger instances. The proposed framework is fully open-source and available at \redacted{\href{https://github.com/cda-tum/mqt-debugger}{https://github.com/cda-tum/mqt-debugger}}.

The remainder of this work is structured as follows: \autoref{sec:background} reviews core concepts and related work in the fields of quantum computing and debugging. \autoref{sec:motivation} then illustrates the considered problem and devises a general idea to mitigate it. Following this, \autoref{sec:proposed-framework} proposes a framework for efficient debugging of quantum programs based on this general idea, highlighting its main features, including the different types of supported assertions and its employed diagnosis methods. \autoref{sec:application} then evaluates the applicability and limitations of this framework for larger problem instances. Finally, \autoref{sec:conclusion} suggests possible future improvements and concludes this work.

\section{Background}\label{sec:background}
This section describes existing work in the area of classical debugging and related techniques for quantum programs.
\subsection{Classical Debugging}

Debugging is a crucial part of the classical development workflow. Interactive debugging tools, such as GDB~\cite{gdb}, are commonly employed by most developers to efficiently find and resolve errors in programs. To make such debugging tools more accessible, the \emph{Debug Adapter Protocol}~\cite{dap} has been proposed to specify a standard for interactive debugging software. 

Since these software tools still require manual work from developers, automated debugging and analysis approaches are another popular area of research that attempt to infer potential problems in software without active user interaction. In many cases, they are based on \emph{static analysis}~\cite{ayewah2008}, examining programs without actively running the code and instead inspecting the individual instructions and their relations on a general level. In contrast, dynamic analysis~\cite{ball1999} focuses on analyzing the properties of programs while executing them.

Automated debugging methods typically attempt to reduce the number of lines that need to be inspected to allow efficient error searches even in large programs. One way to achieve this is through \emph{program slicing}~\cite{weiser1984}, a method used to reduce a program by inspecting its control and data flow. In 1999, Zeller~\cite{zeller1999} introduced \emph{delta debugging}, an efficient method of finding error causes by inspecting changes between a correct original implementation and an update that introduces an error.

The concept of debugging can also be applied to classical circuits. While automated circuit verification is a commonly employed method to test circuits for their \mbox{correctness~\cite{bryant1986,kuehlmann2003,molitor2010}}, several methods have been proposed to automatically find error causes in different types of circuits~\cite{veneris1999, smith2005, ali2005, wille201151, strasnick2021} once verification has discovered the presence of errors.

\subsection{Assertions and Debugging of Quantum Programs}
\label{sec:background-assertions}
Assertions are a powerful tool in the process of debugging and testing software~\cite{rosenblum1995}. They are represented by expressions that can be embedded in software code. During runtime, these expressions are evaluated and, depending on the evaluation results, the execution is aborted, and errors are reported.

In the context of quantum computing, several approaches to define quantum-state assertions for formal verification and runtime testing on real devices have been proposed~\cite{liu2019, haner2020, yu2021, ying2022, bauer-marquart2023, huang2019, li2020, liu2020, liu2021, witharana2023}. When evaluating assertions on quantum devices, however, additional care has to be taken to prevent them from impacting the system's state. Measurements required to extract the state of individual qubits may cause side effects, collapsing parts of the system. 

To circumvent this problem, these approaches propose methods that leave the state of the system unchanged at the cost of increasing the complexity of the circuit---by introducing ancilla qubits~\cite{liu2020,liu2021} or performing projective measurements~\cite{li2020}. These trade-offs increase the depth and width of the corresponding circuits, negatively impacting the algorithms' performance, especially on \emph{Noisy Intermediate-Scale Quantum} (NISQ) devices. Huang et al.~\cite{huang2019} proposed a framework for statistical assertions that measures the probability distribution of the current state and, then, terminates the execution, comparing the measured results with the expected values.

Different types of assertions have been proposed that allow developers to test a program for several different properties. Examples include precise and approximate assertions on the system's full state or its individual amplitudes~\cite{liu2021, liu2020}, assertions testing for entanglement or superposition properties of the \mbox{system \cite{huang2019, witharana2023}}, as well as assertions that require the current state to be contained in the span of a set of state vectors~\cite{yu2021}.

Despite the advantages of these quantum assertions, their deployment still requires manual work from experts in the field. Developers must place assertions manually, which is a time-consuming task that requires a deep understanding of the full algorithm. Moreover, even if assertions detect an issue, identifying the precise location of the underlying cause is not trivial and often requires further manual work from experts to find it. Witharana et al.~\cite{witharana2023} propose \emph{quAssert}, a framework that employs static analysis to extract locations for efficient assertions from a quantum circuit definition. However, this approach requires the existence of a "ground truth" circuit functioning without errors to be employed. Similarly, \mbox{Li et al.~\cite{li2014}} propose a method to compute fitting locations for projective measurements to monitor the correctness of a given operator.  Both of these approaches, however, only assist in the placement of assertions, not in the task of error location.

\section{Motivation and General Idea}\label{sec:motivation}

This section further highlights current shortcomings and challenges in developing and debugging quantum algorithms. Based on this, it then devises a general idea to reduce the required workload and assist developers in the debugging process.

\subsection{Considered Problem}

Even with the help of existing debugging tools for quantum programs, locating error causes remains a challenging task. This process is further complicated by the increased availability of quantum software tools for the automated design of quantum computing applications (such as~\cite{qiskit, bergholm2018pennylane, quetschlich2023problemsolver, dimatteo2023quantum, rovara2024pathfindingframework,volpe2024towards, apak2024ketgpt, quetschlich2023mqtbench}), as well as the ongoing development of quantum programming languages (such as~\cite{seidel2024qrispframeworkcompilablehighlevel,cross2022openqasm,bichsel2020silq,smith2017practicalquantuminstructionset,qsspec2024}). These developments allow users to create quantum programs with more and more lines of code and, hence, lead to a rapidly increasing potential for errors in the corresponding programs. Due to the more complex nature of quantum algorithms compared to their classical counterparts, finding such errors can be deceptively difficult, even in comparatively simple circuits. 

\begin{example}
    Consider the instance of Grover's algorithm~\cite{grover1996fast} with the target state $\left| 111 \right \rangle$ as defined in \autoref{lst:grover-with-errors}, which will act as a running example throughout this paper. After two Grover iterations, one would expect to measure $\left| 111 \right \rangle$ with a high probability. However, this implementation is incorrect. Moreover, even for this considerably small quantum algorithm composed of only 23 lines of code, it is not easy to find the errors at first glance.

    The first error is located in the \texttt{ccz} instruction in Line~2. Clearly, the oracle function, responsible for marking the target state $\left| 111 \right \rangle$ should take all three qubits of the data register \texttt{q} into account, not just \texttt{q1} and \texttt{q2}. To resolve this issue, Line~2 should be replaced by \texttt{cccz q0, q1, q2, flag}. 
    
    The second error is hidden on a deeper level, as it is not caused by an incorrect configuration but by the absence of a gate. For Grover's algorithm to work correctly, the quantum register \texttt{q} has to be put into an equal superposition using Hadamard gates (\texttt{h q;}) before Line~17. Only after these gates are added will the algorithm yield the expected result.
\end{example}

\begin{lstlisting}[language=mqt-qasm,caption=An incorrect implementation of Grover's algorithm using 4 qubits. This instance of the algorithm is supposed to search for the state $\left| 111 \right \rangle$.,label=lst:grover-with-errors]
gate oracle q0, q1, q2, flag {
    ccz q1, q2, flag;
}

gate diffusion q0, q1, q2 {
    h q0; h q1; h q2;
    x q0; x q1; x q2;
    ccz q0, q1, q2;
    x q2; x q1; x q0;
    h q2; h q1; h q0;
}

qreg q[3]; qreg flag[1]; creg c[3];
                             
x flag;

oracle q[0], q[1], q[2], flag;
diffusion q[0], q[1], q[2];

oracle q[0], q[1], q[2], flag;
diffusion q[0], q[1], q[2];

measure q -> c;
\end{lstlisting}

Both of the issues in the above example are frequently occurring problems in the implementation of quantum algorithms. The misconfiguration of controlled quantum gates, e.g., by switching or missing individual controls, can often go unnoticed, especially in larger code bases containing multiple such gates. Missing individual gates, on the other hand, may also lead to larger issues. This is even more so the case, when the consequences are only encountered in different parts of the code, such as in custom gate definitions. These custom gates often require certain pre-conditions to function---effectively forcing developers to search the entire code base for possible sources of errors. 

As a result, manually inspecting quantum algorithms for possible errors is a difficult and time-consuming task that requires an exact mathematical understanding of the entire system at each point in time. Furthermore, as the state space grows exponentially with respect to the number of qubits, the limitations of manual processing are easily reached, even for small examples. While it may still be possible to check the above example by hand, adding just four more qubits increases the number of amplitudes that have to be considered to 256. Therefore, efficient automated tools that support developers in the search for error causes are crucial for real-world applications.

\subsection{General Idea}

To assist developers in this process, this work introduces a framework to aid in the identification and resolution of errors in quantum programs. Its methods are based on the \emph{classical simulation} of quantum circuits. Compared to execution on physical quantum devices, this allows it to avoid physical limitations of quantum computing, such as the \emph{no-cloning theorem}~\cite{wootters1982} and provide continuous access to the full quantum state of the system during each step of the execution.

In order to test the current state of the system during execution, the proposed framework employs \emph{assertions}. These assertions are then examined during the simulation process, and their results are used to perform automated diagnosis on the program.

While the placement of assertions still needs to be conducted manually, the proposed framework supports developers by providing diagnosis methods once an assertion has failed via automatic analysis passes. By investigating the previous interactions of qubits involved in a failed assertion and analyzing the
state of the system during the simulation, the proposed solution \emph{automatically} infers possible causes for the error. While finding the exact location of errors may not always be feasible or even
possible, the framework takes advantage of the reported results to minimize the number of instructions that need to be manually inspected by the developer. To demonstrate the benefits of this approach, the ideas are illustrated using the running example again:

\begin{tikzpicture}[overlay]
    \redactedInstead{\draw[black] (-0.45cm,-0.2cm) rectangle (8.7cm,-8.95cm);}{\draw[black] (-0.45cm,-0.2cm) rectangle (8.7cm,-9.22cm);}
\end{tikzpicture}

\begin{lstlisting}[language=mqt-qasm,caption={The incorrect implementation of Grover's algorithm from \autoref{lst:grover-with-errors}, including new quantum-state \emph{assertions} and suggested problem causes. Here, red wavy underlines indicate possible error locations, and yellow lightning symbols highlight failing assertions.},label=lst:grover-with-assertions,frame=none]
gate oracle q0, q1, q2, flag {
    (*@\lightning@*)assert-sup q0, q1, q2;(*@\graybox{Qubits are not in a superposition. Are you missing a gate before Line 19?}@*)
    (*@\squigglyunderline{\textbf{ccz} q1, q2, flag;}@*) 
    (*@\lightning@*)assert-ent q0, q1, q2;(*@\graybox{Qubits are not entanlged. Is there a control qubit missing on Line 3?}@*)
}

gate diffusion q0, q1, q2 {
    h q0; h q1; h q2;
    x q0; x q1; x q2;
    ccz q0, q1, q2;
    x q2; x q1; x q0;
    h q2; h q1; h q0;
}

qreg q[3]; qreg flag[1]; creg c[3];
                                
x flag;
(*@\squigglyunderline{\hspace{3cm} }@*)
oracle q[0], q[1], q[2], flag;
diffusion q[0], q[1], q[2];
(*@\lightning@*)assert-eq 0.8, q { 0, 0, 0, 0, 0, 0, 0, 1 }

oracle q[0], q[1], q[2], flag;
diffusion q[0], q[1], q[2];
(*@\lightning@*)assert-eq 0.9, q { 0, 0, 0, 0, 0, 0, 0, 1 } 

measure q -> c;
\end{lstlisting}

\begin{example}
Returning to the instance of Grover's algorithm considered in \autoref{lst:grover-with-errors}, we place assertions on \mbox{Lines 2, 4, 21, and 25,} as illustrated in \autoref{lst:grover-with-assertions}. The assertion on Line~2 is used as a pre-condition to enforce that the given qubits are in a superposition state. This ensures that the oracle is not used on invalid input states. The assertion on Line~4, on the other hand, checks whether the oracle function itself acts as expected. Finally, the assertions on Lines~21~and~25 test the functionality of the overall circuit.

Simulating the algorithm while evaluating these assertions and employing the corresponding possible diagnosis methods (detailed more thoroughly later in \autoref{sec:proposed-framework}) reveals a potential error in Line~3. In particular, static analysis reveals the absence of interactions between qubit \texttt{q0} and the remainder of the data register \texttt{q}, which prevents entanglement. Furthermore, the failure of the pre-condition assertion on Line~2 and subsequent analysis at runtime indicate the possibility of a missing gate before Line~19. This way, instead of 23 lines, the developer only has to inspect 4 lines to resolve all errors (all of them explicitly highlighted by the proposed framework).
\end{example}

In the following, the proposed approach is described in more detail.

\section{Proposed Framework}\label{sec:proposed-framework}

In this section, we describe the framework introduced above. This includes a more detailed definition of the required inputs, as well as all the analysis methods employed for diagnosis, including a \emph{cone of influence analysis}, an \emph{interaction analysis}, and a \emph{control value analysis}. For additional information, we refer to its open-source implementation, available at \redacted{\href{https://github.com/cda-tum/mqt-debugger}{https://github.com/cda-tum/mqt-debugger}}.

\subsection{Input Specification}

The proposed framework assumes the specification of a quantum program (currently based on the \mbox{OpenQASM}~\cite{cross2017} standard), which includes manually placed \emph{assertions} to test for its correctness. The assertions are assumed to be expressed as pre-defined OpenQASM-style instructions. At the moment, the proposed framework supports three types of assertions on quantum states: 
\begin{itemize}
    \item \emph{entanglement-assertions} (\texttt{assert-ent}) test, whether a given set of qubits is fully entangled,
    \item \emph{superposition-assertions} (\texttt{assert-sup}) test, whether the given set of qubits is in a superposition state, and
    \item \emph{equality-assertions} (\texttt{assert-eq}) test, whether the quantum state of a system is equal to the provided reference state (up to some similarity margin).
\end{itemize}

\begin{example}
The quantum program in \autoref{lst:grover-with-assertions} includes four assertions on \mbox{Lines 2, 4, 21, and 25}. The superposition assertion on Line~2 enforces that the three qubits of the data register are in some superposition state. At this point, we expect them to be in the state $\frac{1}{2\sqrt{2}}\sum_{x=0}^7 |x\rangle$, which clearly constitutes a superposition. However, due to the incorrect initialization, their state will instead be $|000\rangle$ at the first execution, causing the assertion to fail.

Line~4, on the other hand, requires that the data register is fully entangled, i.e., non-separable. However, as the data register is still in the state $|000\rangle$ at its first execution, the assertion will fail since it can trivially be separated into \mbox{$|0\rangle \otimes |0\rangle \otimes |0\rangle$}.

Finally, the equality assertions on \mbox{Lines 21 and 25} both test whether the data register \texttt{q} equals the target state $|111\rangle$. The assertion on Line~21 requires a similarity of 0.8, while the assertion on Line~25 requires a similarity of 0.9. With more iterations of the Grover circuit, we expect \texttt{q} to converge toward the target state $|111\rangle$. However, due to the errors in the program, Grover's algorithm is not performed correctly and both assertions fail.
\end{example}

The failures of any of these assertions provide valuable information on the underlying errors in the code. To leverage this information, the proposed framework provides a set of diagnosis methods, that can be used to locate the corresponding error causes. They are described in the following.

\subsection{Cone of Influence Analysis}
By statically analyzing the data dependencies of a failed assertion, the program's instructions can be partitioned into two sets, based on whether an instruction influences any of the qubits contained in the failed assertion (also called the \emph{cone of influence}) or not. These results can then be employed in several ways: By highlighting relevant instructions in an integrated development environment, users can manually skim through the smaller subset of instructions, searching for incorrect instructions inside the cone of influence, or for instructions that are missing from it. Furthermore, this cone of influence analysis can also be taken as the basis for further automated diagnosis methods.
\begin{example}
Returning to \autoref{lst:grover-with-assertions}, each of the failing assertions acts solely on the data register \texttt{q}. The cone of influence analysis will highlight all lines above the corresponding assertion that directly or indirectly affect \texttt{q}. For the assertion on Line~2, this includes \mbox{Lines 1, 15, and 19}. This allows the developer, as well as the remaining analysis steps, to investigate just these four lines for errors, or, alternatively, conclude that a crucial gate may be missing.

As the execution progresses, the number of gates inside the cone of influence tends to increase. The assertion on Line~4 includes \mbox{Lines 3 and 17} in its cone of influence, while the cone of influence for the assertion on Line~21 includes every instruction from Line~1 to Line 20. The proposed framework highlights the lines included in the cone of influence automatically, aiding the developer in inspecting the code.
\end{example}

The applicability of this method strongly depends on the nature of a given quantum program. Algorithms that require a large number of interactions between qubits can easily cause a majority of the instructions to be included in a given cone of influence, especially at later steps of execution. However, as discussed later in \autoref{sec:application}, certain types of algorithms can profit from this approach even at larger sizes, and any improvement of performance greatly increases debugging capabilities. 

\subsection{Interaction Analysis}
For two qubits to be entangled with each other, they must interact directly or indirectly in some previous instructions. Interaction analysis searches for qubits used in an entanglement assertion that do not interact with each other in any previous instruction, reporting them to the user. While this method can only be employed specifically for entanglement assertions, it can be a helpful tool specifically for the error type of misconfigured controlled gates, easily detecting gates that are missing individual controls.

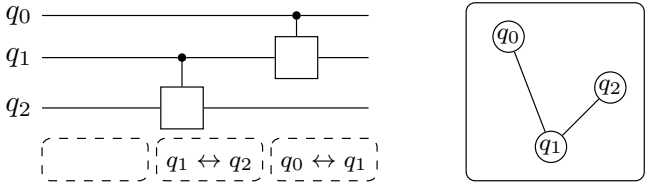
\begin{figure}
    \centering
       \resizebox{1\linewidth}{!}{
            \begin{tikzpicture}
                \begin{yquant}	
                    qubit {$q_0$} q[+1];
                    qubit {$q_1$} q[+1];
                    qubit {$q_2$} q[+1];

                    hspace {1.3cm} -;
                                        
                    box {\hspace{0.3cm}} q[2] | q[1];
                    hspace {0.75cm} -;

                    box {\hspace{0.3cm}} q[1] | q[0];
                    hspace {0.5cm} -;                    
                \end{yquant}
                
                \draw[rounded corners=3pt, dashed] (0, -1.6) rectangle (1.25, -2.1);
                \node at (0.625, -1.9) {};

                \draw[rounded corners=3pt, dashed] (1.35, -1.6) rectangle (2.6, -2.1);
                \node at (1.975, -1.9) {\footnotesize $q_1 \leftrightarrow q_2$};

                \draw[rounded corners=3pt, dashed] (2.7, -1.6) rectangle (3.95, -2.1);
                \node at (3.325, -1.9) {\footnotesize $q_0 \leftrightarrow q_1$};

                \draw[rounded corners=3pt] (5, 0) rectangle (7.1, -2.1);
                \node[draw, circle, inner sep=0.4pt] (q0d) at (5.5, -0.4) {\footnotesize $q_0$};
                \node[draw, circle, inner sep=0.4pt] (q1d) at (6, -1.7) {\footnotesize $q_1$};
                \node[draw, circle, inner sep=0.4pt] (q2d) at (6.7, -1) {\footnotesize $q_2$};
                \draw (q1d) -- (q2d);
                \draw (q0d) -- (q1d);

            \end{tikzpicture}
        }
    \caption{The process of constructing interaction graphs by collecting gates acting on multiple qubits. Two qubits are said to interact if there exists a path between them on the interaction graph.}
    \label{fig:interactions}
    \vspace{-0.5cm}
\end{figure}

For this purpose, the interaction of two qubits $x$ and $y$ is defined recursively, such that $x$ interacts with $y$ if and only if there exists a gate that acts on both $x$ and $y$, or if there exists a gate acting on $x$ and some qubit $z$, where $z$ interacts with $y$. To evaluate this, we perform a breadth-first search on the instructions included in the failed assertion's cone of influence, constructing an interaction graph for the different qubits, as illustrated in \autoref{fig:interactions}. After the first controlled gate, the interaction between $q_1$ and $q_2$ is added to the interaction graph. Then, the controlled gate between $q_0$ and $q_1$ adds an additional connection to the graph. As there now exists a path from $q_0$ to $q_2$, the qubits are proven to interact with each other.

\begin{example}
After encountering the failing entanglement assertion on Line~4 in \autoref{lst:grover-with-assertions}, the proposed framework performs an interaction analysis between the three data qubits. Iterating over their cone of influence, it finds an interaction between \texttt{q1} and \texttt{q2} in Line~3, but no interactions between \texttt{q0} and other data qubits. The full interaction graph, therefore, consists of only one edge between \texttt{q1} and \texttt{q2}, similarly to the state after the first gate has been considered in \autoref{fig:interactions}. As no path exists between \texttt{q0} and the other qubits, this approach concludes that a controlled gate interacting with \texttt{q0} might be missing, or that Line~3 may be missing a control qubit.
\end{example}

As deep circuits commonly re-use qubits and this definition of interaction does not allow for interactions to be canceled without resetting the corresponding qubits, this diagnosis method is once again only partially applicable to larger circuits. However, especially during early steps of the execution, this method still provides valuable assistance to the developer. Especially in programs that employ several custom gate definitions, it shields developers from being required to follow the execution through all the required jumps and points them easily toward possible errors.

\subsection{Control Value Analysis}
During execution, the proposed framework keeps track of all controlled gates that can only perform the identity operation. This happens if any of the control qubits used in a controlled gate is always zero. As the proposed framework has access to the state vector during the entire execution process, this dynamic analysis task can be applied easily. When encountering a controlled gate for which all control qubits are zero, the framework marks them as potential problems. Should a later execution of the same instruction no longer have this problem, then the mark will be removed. If a failing assertion is encountered, it will investigate whether any of the instructions within its cone of influence have been marked this way, and report those instructions. Once again, this method can easily help developers in finding misconfigured controlled gates. In particular, cases in which control and target have been swapped, as well as missing instruction during state preparation can often be found by this analysis method.
\begin{example}
    When the \texttt{ccz} gate in Line~3 ist first encountered, it will note that the control qubits are in state $|00\rangle$ at the time and mark this gate as a potential error candidate. Once any assertion that includes Line~3 in its cone of influence is encountered, the automatically performed control value analysis will highlight that the controls in Line~3 are always zero, and infer that a gate may be missing before Line~3 in the cone of influence.
    When the \texttt{oracle} gate is called a second time, the data register will have been modified by previous instructions. Because of this, the second execution of Line~3 will remove the mark from the instruction, and, consequently, the control value analysis employed after the failing assertion on Line~25 will not highlight this issue again.
\end{example}

This analysis method can also be employed efficiently on deep circuits. As controlled gates typically only act on a small number of qubits, the number of variables to be investigated for each gate remains limited. Furthermore, due to the crucial role controlled gates play in quantum circuits, this analysis method may be employed to find error causes for all types of assertions and can find applications in all types of quantum programs.

\redactedInstead{\vspace{-0.3cm}}{\vspace{-0.3cm}}
\section{Application and Evaluation}
\label{sec:application}

The methods proposed above have been implemented and made available as an open-source framework at~\redacted{\href{https://github.com/cda-tum/mqt-debugger}{https://github.com/cda-tum/mqt-debugger}}. The running example and discussions above already demonstrated its applicability and usefulness. However, to confirm the corresponding benefits also on larger instances, further tests and evaluations have been conducted.
The corresponding results are summarized in the following.

Due to page limitation, we hereby focus on two representative instances (the availability as an open-source framework easily allows to try and test further instances), namely the programs \texttt{dj-130}, implementing the \mbox{Deutsch-Jozsa} algorithm~\cite{deutsch1992} on 130 qubits using a total of 391 instructions, as well as \texttt{random-13}, representing a random circuit on 13~qubits that uses 432 instructions. Both of these circuits were generated using the benchmark library MQT Bench~\cite{quetschlich2023mqtbench} and were selected to properly illustrate the benefits (but also limitations) of the proposed methods. For all considered evaluations, the analysis methods were performed in real-time.
Even as the program size increased, the algorithms efficiently performed their analysis tasks, with the primary performance bottleneck remaining the classical quantum circuit simulation.
\redactedInstead{To this end, we have employed a dedicated simulator backend using decision diagrams as a data structure~\cite{zulehner2017, willeToolsQuantumComputing2022}.}{}

In the following, we discuss the performance of the framework and, in particular, each analysis method individually, highlighting their advantages as well as possible reasons for their limitations in more detail.

\subsubsection{Cone of Influence Analysis}

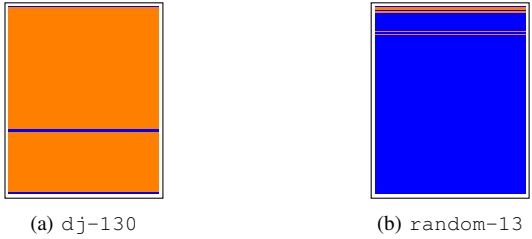
\begin{figure}
    \centering
    \begin{subfigure}{0.45\columnwidth}
        \centering
        \begin{tikzpicture}[scale=0.025]

            \draw (-2,-2) rectangle (82,102);
            \fill[orange] (0,0) rectangle (80,0.5);
            \fill[blue]  (0,0.5) rectangle (80,1.25);
            \fill[orange] (0,1.25) rectangle (80,33.5);
            \fill[blue]  (0,33.5) rectangle (80,35.25);
            \fill[orange] (0,35.25) rectangle (80,99.75);
            \fill[blue]  (0,99.75) rectangle (80,100);
            
        \end{tikzpicture}
        \caption{\texttt{dj-130}}
    \end{subfigure}
    \hfill
    \begin{subfigure}{0.45\columnwidth}
        \centering
        \begin{tikzpicture}[scale=0.025]

            \draw (-2,-2) rectangle (82,102);  %
            \fill[orange] (0,0) rectangle (80,0.5);
            \fill[blue]  (0,0.5) rectangle (80,85.0);
            \fill[orange] (0,85) rectangle (80,85.25);
            \fill[blue]  (0,85.25) rectangle (80,86.75);
            \fill[orange] (0,86.75) rectangle (80,87);
            \fill[blue]  (0,87) rectangle (80,89.5);
            \fill[orange] (0,89.5) rectangle (80,90);
            \fill[blue]  (0,90) rectangle (80,96.5);
            \fill[orange] (0,96.5) rectangle (80,97.5);
            \fill[blue]  (0,97.5) rectangle (80,98.5);
            \fill[orange] (0,98.5) rectangle (80,99.75);
            \fill[blue]  (0,99.75) rectangle (80,100);
            
        \end{tikzpicture}
        \caption{\texttt{random-13}}
    \end{subfigure}
    \caption{The partitions inside (blue) and outside (orange) the cone of influence for an assertion over three qubits at the end of the evaluation programs.\redactedInstead{\vspace{-0.5cm}}{\vspace{-0.5cm}}}
    \label{fig:cone-of-influence}
\end{figure}

For both programs, the proposed framework performed a cone of influence analysis over an equality assertion involving three qubits at the end of execution.
\autoref{fig:cone-of-influence} illustrates the partitions of the input program returned by this analysis.
Each of the boxes represents the  code of the corresponding program.
Orange sections represent lines of code that are outside of the cone of influence, whereas blue sections represent lines of code inside the cone of influence.

The results confirm the expectation that dedicated quantum algorithms exhibit significant local structure, which is advantageous for this type of analysis.
Out of the 391 lines of the \texttt{dj-130} program, only 11 are included in the cone of influence, reducing the number of lines to be considered for debugging by \emph{more than 97\%}.

However, the situation is different for the \texttt{random-13} program.
Given that deep random circuits typically involve pairwise interactions between all qubits, the cone of influence grows significantly.
As a result, out of the 432 lines of \texttt{random-13}, 416 are included in the cone of influence, merely reducing the number of lines to be considered by 4\%.

Overall, the cone of influence analysis supports developers in debugging quantum programs by reducing the number of lines to be considered for locating errors.

\subsubsection{Interaction Analysis}

\begin{figure}
    \centering
    \begin{subfigure}{0.4\columnwidth}
        \centering
        \begin{tikzpicture}[scale=0.37]
            \tikzstyle{every node}=[draw, circle, minimum size=4mm, inner sep=0mm]
            
            \node (129) at (0, 0) {129};
            
            \foreach \i/\name in {0/0, 1/1, 2/2, 3/3, 4/4, 5/5, 6/6, 7/7, 8/8, 9/9, 
                                  10/10, 11/11, 12/12, 13/13, 14/14, 15/15, 16/16, 17/17,
                                  18/18, 19/19, 20/20, 21/21, 22/22, 23/23, 24/24, 25/25,
                                  26/26, 27/27, 28/28, 29/29, 30/30, 31/31}
                \node (\name) at ({360/32 * \i}:5.5) {\footnotesize \name};
            
            \foreach \name in {0, 1, 2, 3, 4, 5, 6, 7, 8, 9, 10, 11, 12, 13, 14, 15, 
                               16, 17, 18, 19, 20, 21, 22, 23, 24, 25, 26, 27, 28, 
                               29, 30, 31}
                \draw (129) -- (\name);
        \end{tikzpicture}
        \caption{\texttt{dj-130}}
    \end{subfigure}
    \hfill
    \begin{subfigure}{0.45\columnwidth}
        \centering
        \raisebox{1cm}{
        \begin{tikzpicture}[scale=0.5, every node/.style={draw, circle, minimum size=4mm, inner sep=0mm}]
            \node (0) at (0, 0) {0};
            \node (1) at (2, 1) {1};
            \node (2) at (3, 0) {2};
            \node (3) at (3.5, 1) {3};
            \node (4) at (2, -2) {4};
            \node (5) at (-1.5, -2) {5};
            \node (6) at (1, 0) {6};
            \node (7) at (-1.5, 0) {7};
            \node (8) at (3, 2) {8};
            \node (9) at (0, 2) {9};
            \node (10) at (3.5, -1) {10};
            \node (11) at (-0.5, -1) {11};
            \node (12) at (-1.5, 2) {12};
            \draw (0) -- (9);
            \draw (8) -- (9);
            \draw (7) -- (9);
            \draw (1) -- (8);
            \draw (1) -- (6);
            \draw (9) -- (12);
            \draw (5) -- (7);
            \draw (5) -- (11);
            \draw (7) -- (11);
            \draw (3) -- (8);
            \draw (0) -- (4);
            \draw (2) -- (4);
            \draw (2) -- (10);
            \draw (4) -- (10);
        \end{tikzpicture}
        }
        \caption{\texttt{random-13}}
    \end{subfigure}
    \vspace*{-3mm}
    \caption{The interaction graphs for both evaluation programs, as computed by the proposed framework after 50\% of their operations have been applied. For~\texttt{dj-130}, qubits 32 to 128 have not yet interacted with any other qubits.}
    \label{fig:interaction-application}
    \redactedInstead{\vspace{-0.5cm}}{\vspace{-0.5cm}}
\end{figure}
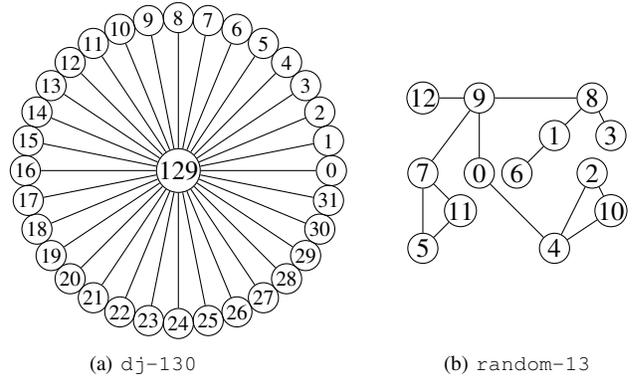

Interaction analysis was performed on both programs to understand the connectivity and interaction patterns among qubits.
Performing this analysis at the end of the programs yields a strongly connected interaction graph in both cases, as all qubits either directly or indirectly interact with other qubits at some point.
However, typically, this analysis is conducted on failing entanglement assertions placed at intermediate positions in the code to identify whether qubits that should be entangled are indeed interacting.
Hence, \autoref{fig:interaction-application} shows the interaction graphs for both programs computed after half of their instructions have been processed.

For the \texttt{dj-130} program, the interaction graph exhibited a simple star topology, connecting only 33 of the 130 qubits.
This simplified structure makes it easier to identify expected interactions and potential issues.
For example, the failure of an assertion requiring qubits 128 and 129 to be entangled can be attributed to the missing interaction of the two qubits in the interaction graph.
In contrast, the \texttt{random-13} program also showed a strongly connected interaction graph halfway through its execution, reflecting the complexity and randomness of the circuit.
In these cases, the analysis remains inconclusive for identifying possible causes of failing assertions.

Overall, the interaction analysis supports developers by allowing them to identify potentially missing interactions between qubits, such as missing controls or entirely missing gates.

\subsubsection{Control Value Analysis}

Control value analysis was conducted to evaluate the framework's performance on large inputs.
Given the number of qubits in the \texttt{dj-130} program, accessing the state vector regularly for this analysis is impractical in this case.
Therefore, we focused on the \texttt{random-13} program, modifying a single, randomly chosen instruction 100 times and running the analysis to determine how often it provides a suggestion that allows the developer to identify the error.
If the randomly chosen instruction was a controlled gate, we flipped the control and target qubits.
If it was another type of instruction, it was removed.

The control value analysis managed to correctly suggest error locations in 22 out of 100 test cases.
In particular, it triggered in 31\% of the test instances when an instruction was removed, while it only triggered in 10\% of the instances when a controlled gate was flipped.
This discrepancy likely arises because removing gates, especially early in the program, can disrupt important state preparation steps, leaving qubits in the $|0\rangle$ state longer and allowing the analysis to trigger more frequently.
Additionally, the analysis showed a 75.9\% success rate in identifying errors introduced before Line 100.
Errors introduced after Line 100 were not correctly identified in any test case, as the system's state became too disordered due to the random nature of the circuit.

Overall, the control value analysis is especially helpful for identifying errors early in the program, pointing developers toward their possible causes.

\vfill
\redactedInstead{}{\vspace*{-1mm}}
\section{Conclusion}\label{sec:conclusion}
\redactedInstead{}{\vspace*{-1mm}}

In this work, we proposed a framework that allows developers to efficiently debug quantum programs.
Given a set of assertions, it simulates provided programs and checks the assertions for correctness.
Upon encountering an unexpected state, it uses several automated analysis methods to diagnose the problem and find the underlying causes.
Evaluations showed that, in many cases, these methods can be employed to efficiently diagnose problems in programs consisting of several hundred instructions.
Future work involves devising further analysis methods and extending the workflow by using analysis results to suggest better locations for assertions when diagnosis fails.
The proposed debugging framework is available as an open-source implementation at \redacted{\href{https://github.com/cda-tum/mqt-debugger}{https://github.com/cda-tum/mqt-debugger}}.

\vfill

\redactedInstead{
\section*{Acknowledgments}
This work received funding from the European Research Council (ERC) under the European Union’s Horizon 2020 research and innovation program (grant agreement No. 101001318), was part of the Munich Quantum Valley, which is supported by the Bavarian state government with funds from the Hightech Agenda Bayern Plus, and has been supported by the BMWK on the basis of a decision by the German Bundestag through project QuaST, as well as by the BMK, BMDW, and the State of Upper Austria in the frame of the COMET program (managed by the FFG).
}
{}

\clearpage

\printbibliography

\end{document}